\documentclass[3p,times,twocolumn]{elsarticle}
\biboptions{comma,sort&compress}
 
\usepackage{graphicx}
\usepackage{subfig}
\usepackage{here}
\usepackage{ecrc}
\usepackage{float}
\usepackage{multirow}
\usepackage{color}
\usepackage{ulem}

\usepackage{amsmath}
\usepackage{amssymb}

\newcommand{\vev}[1]{\langle\Omega|#1|\Omega\rangle}
\newcommand{\rom}[1]{\uppercase\expandafter{\romannumeral #1\relax}}
\newcommand{\hypgeom}[2]{{}_{#1}F_{#2}}
\newcommand{\lsr}{\mathcal{R}_{0}}
\newcommand{\double}[2]{(#1,\,#2)}
\newcommand{\glueFourD}{\big\langle \alpha G^{2} \big\rangle}
\newcommand{\glueSixD}{\big\langle g^{3} G^{3} \big\rangle}

\volume{00}

\firstpage{1}

\journalname{Nuclear and Particle Physics Proceedings}

\runauth{}

\jid{nppp}

\jnltitlelogo{Nuclear and Particle Physics Proceedings}

\usepackage{amssymb}

\usepackage[figuresright]{rotating}

\usepackage{ulem}

\usepackage{xcolor}

\begin{document}

\begin{frontmatter}

\title{ Heavy Quarkonium ($1^{--}$) Meson-Hybrid Mixing from QCD Sum Rules
 $^*$}
 
 
 \cortext[cor0]{Talk given at 20th International Conference on Quantum Chromodynamics (QCD 17),  3 july - 7 july 2017, Montpellier - FR}
 
 \author[label1]{A. Palameta\fnref{fn1}}
   \fntext[fn1]{Speaker, Corresponding author.}
\ead{a.palameta@usask.ca}
\address[label1]{Department of Physics and Engineering Physics, University of Saskatchewan, Saskatoon, SK, S7N 5E2, Canada}

 \author[label1]{J. Ho}
\ead{josha@joshaho.com}

 \author[label2]{D. Harnett}
    \ead{derek.harnett@ufv.ca}
\address[label2]{Department of Physics, University of the Fraser Valley, Abbotsford, BC, V2S 7M8, Canada}

 \author[label1]{T. G. Steele}
\ead{tom.steele@usask.ca}


\pagestyle{myheadings}
\markright{ }

\begin{abstract}
We use QCD Laplace sum-rules to explore mixing between conventional mesons and hybrids in the heavy quarkonium vector $J^{PC}\!=\!1^{--}$ channel. Our cross-correlator includes perturbation theory and contributions proportional to the four-dimensional and six-dimensional gluon condensates. 
We input experimentally determined charmonium and bottomonium hadron masses into both single and multi-resonance models in order to test them for conventional meson and hybrid components.
In the charmonium sector we find evidence for meson-hybrid mixing in the $J/\psi$ and a $\approx4.3$~GeV resonance. In the bottomonium sector, we find that the $\Upsilon(1S)$, $\Upsilon(2S)$, $\Upsilon(3S)$, and $\Upsilon(4S)$ all exhibit mixing.
\end{abstract}


\end{frontmatter}

\section{Introduction}
Hybrids are hadrons containing a quark-antiquark pair as well as an explicit gluon. Hybrids are color singlets and so are allowed by QCD despite not being included in the constituent quark model. So far, hybrids have not been conclusively identified experimentally. 

Exotic quantum numbers ($J^{PC}$) are those inaccessible to conventional 
mesons (i.e., $\{0^{--},\,0^{+-},\,1^{-+},\ldots\}$) whereas non-exotic quantum numbers are those accessible to conventional mesons. All quantum numbers are accessible to hybrids.  Hybrids with non-exotic quantum numbers are expected to mix with conventional mesons resulting in conventional meson-hybrid superpositions.

The discovery of the XYZ resonances~\cite{Brambilla:2010cs,Eidelman:2012vu,BESIII:2016adj}, 
many of which defy a conventional quark model interpretation~\cite{BarnesCloseSwanson1995}, has sparked much interest 
in the search for outside-the-quark-model hadrons (including hybrids)
within the heavy quarkonium sectors. 

QCD Laplace sum rules (LSRs)~\cite{Shifman:1978bx,Shifman:1978by,Reinders:1984sr,narisonbook:2004} 
have been used to explore mixing in several hadronic systems~\cite{Narison:1984bv,Harnett:2008cw,Chen:2013pya}. Here we study mixing between conventional mesons and hybrids in heavy quarkonium ($c\overline{c}$ and $b\overline{b}$) in the non-exotic vector $J^{PC}\!=\!1^{--}$ channel. We calculate the cross-correlator between a conventional meson current~(\ref{CurMes}) and a hybrid current~(\ref{CurHyb}) within the operator product expansion~(OPE).
We include leading-order (LO) $g_s$ contributions from perturbation theory, the dimension-four (4d) gluon condensate, and the 6d gluon condensate.
LSRs are used to analyze single and multi-resonance models of the experimentally determined vector $c\overline{c}$ and $b\overline{b}$ hadron spectra. 
Measured resonance masses are used as inputs and mixing parameters, indicators of meson-hybrid mixing, are extracted as best fit parameters. 

We find that multi-resonance models containing excited states as well as a ground state lead to much better fits between theory and experiment
as compared to single resonance models.
LSRs generally suppress contributions from excited states.
We show explicitly that in these systems excited resonances make significant contributions to 
the LSRs. This is because their coupling to the currents is sufficient to overcome the exponential suppression.
For vector charmonium, we find that conventional meson-hybrid mixing is well-described by a two-resonance model containing the $J/\psi$ and a $\approx4.3$~GeV state. 
For vector bottomonium, we find evidence of mixing in all of the
$\Upsilon(1S)$, $\Upsilon(2S)$, $\Upsilon(3S)$, and $\Upsilon(4S)$ resonances.

\section{QCD Cross-Correlator Calculation} 

We consider the cross-correlator
\begin{align}
  \Pi_{\mu\nu}(q) &= i\!\int d^{4}\!x \;e^{iq\cdot x} 
    \vev{\tau\, j_{\mu}^{(\text{m})}(x)\; j_{\nu}^{(\text{h})}(0)}
    \label{CorFn}\\
   &=\left(\frac{q_{\mu}q_{\nu}}{q^2}-g_{\mu\nu}\right)\Pi(q^2)
     \label{CorFnProj}
\end{align}
between a conventional meson current 
\begin{equation}
   j_{\mu}^{(\mathrm{m})} = \overline{Q}\gamma_{\mu} Q
    \label{CurMes}
\end{equation}
and a hybrid current~\cite{GovaertsReindersWeyers1985}
\begin{equation}
   j_{\nu}^{(\mathrm{h})}=  \frac{g_{s}}{2}\overline{Q}\gamma^{\rho}\gamma_{5}\lambda^{a} \, \bigg( \frac{1}{2}\epsilon_{\nu\rho\omega\zeta}G^{a}_{\omega\zeta} \bigg) \, Q.
     \label{CurHyb}
\end{equation}
The function $\Pi$ in~(\ref{CorFnProj}) probes $1^{--}$ states.

\begin{figure}
\begin{tabular}{cc}
\multicolumn{2}{c}{\includegraphics[width=35mm]{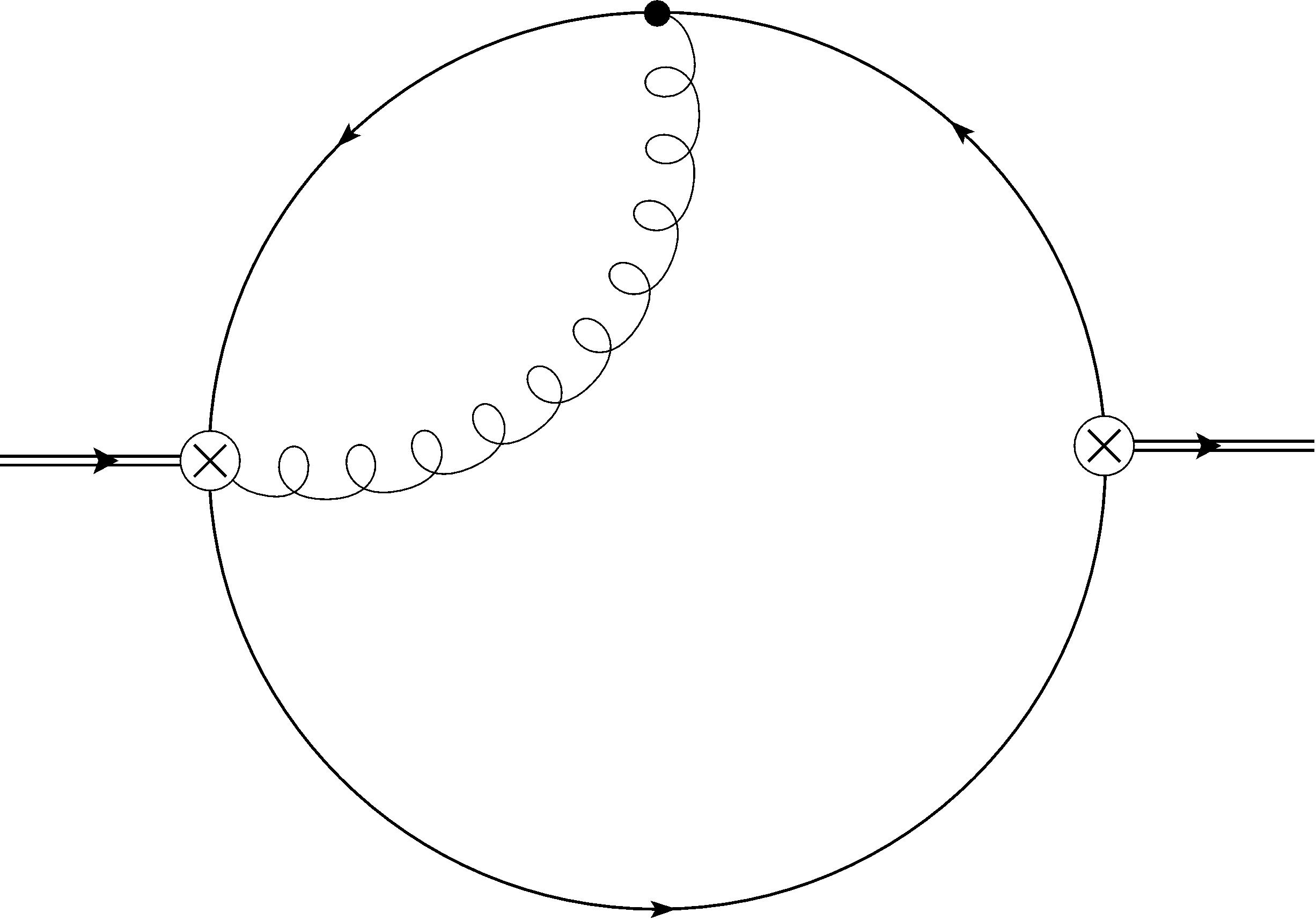} }\\
\multicolumn{2}{c}{Diagram \rom{1}} \\[-3pt]
  \includegraphics[width=35mm]{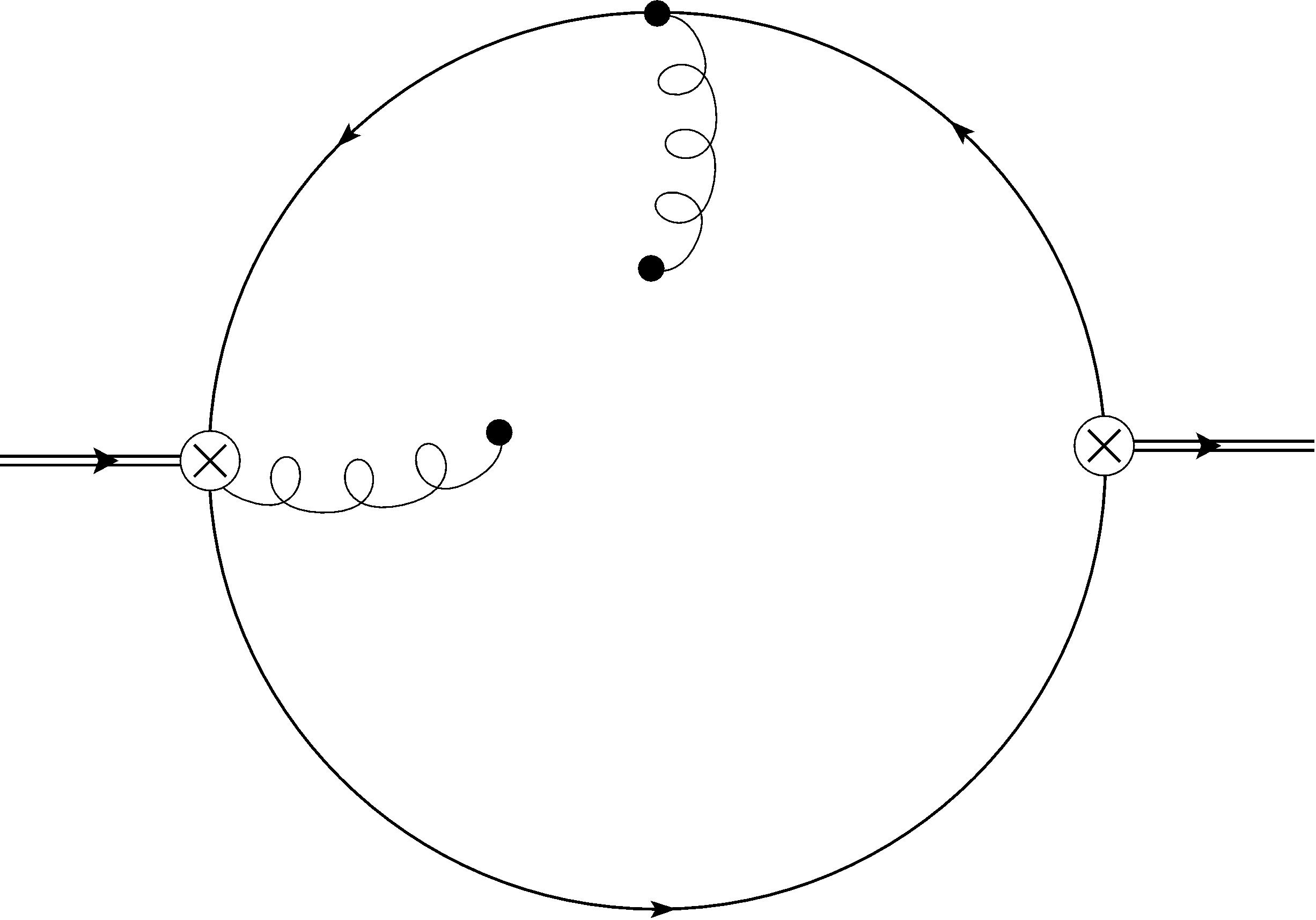} &   \includegraphics[width=35mm]{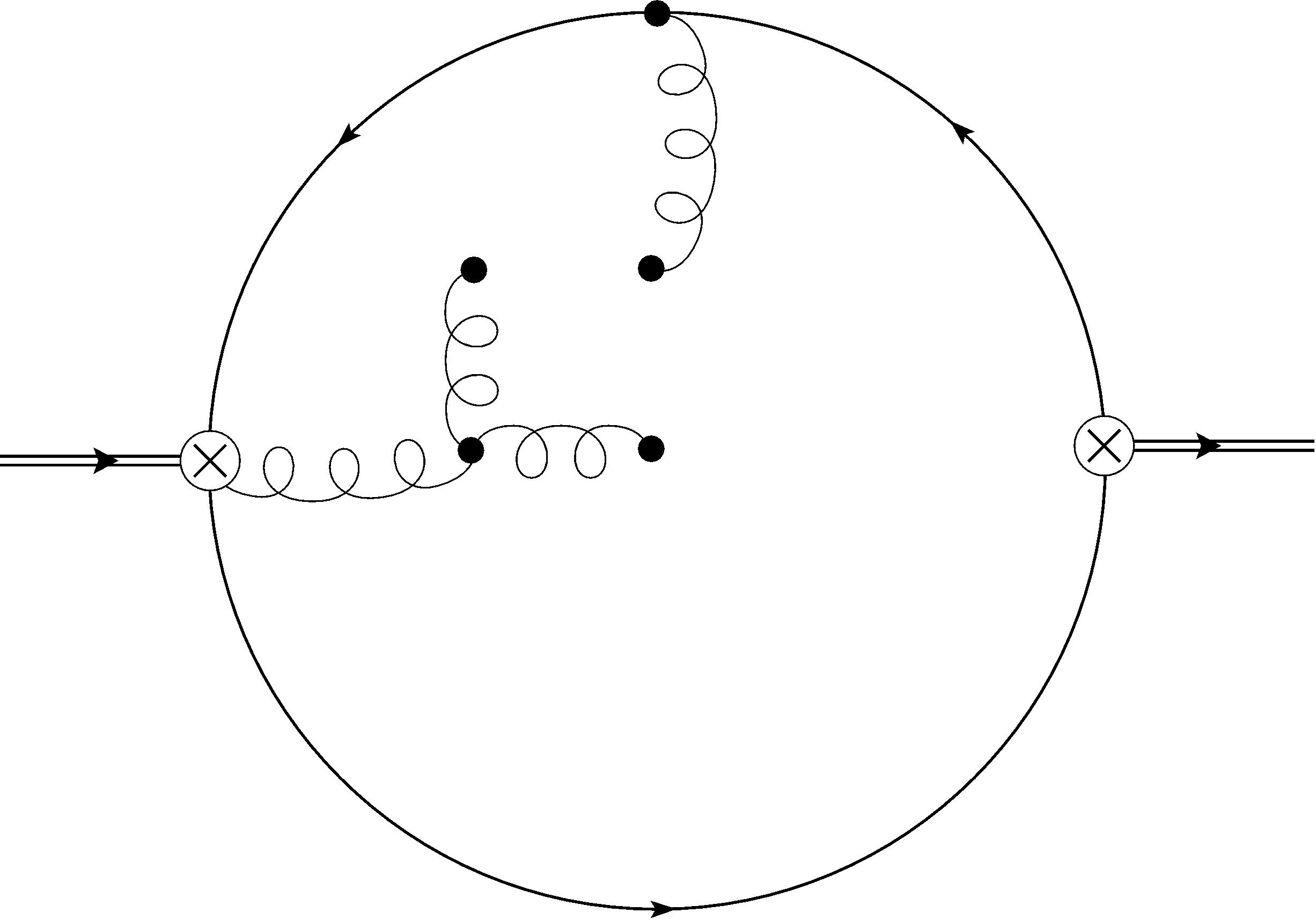} \\
Diagram \rom{2} & Diagram \rom{3} \\[6pt]
 \includegraphics[width=35mm]{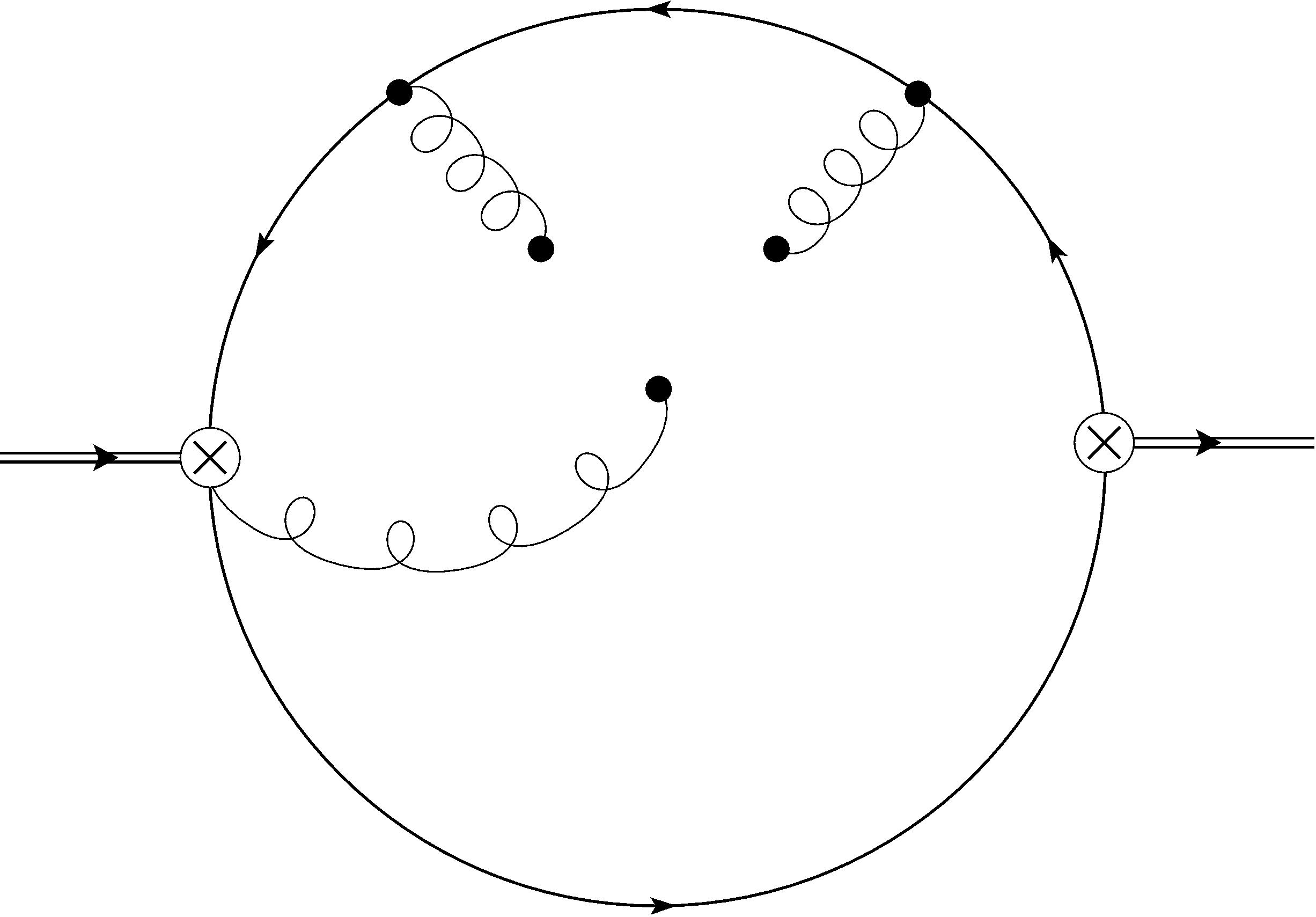} &   \includegraphics[width=35mm]{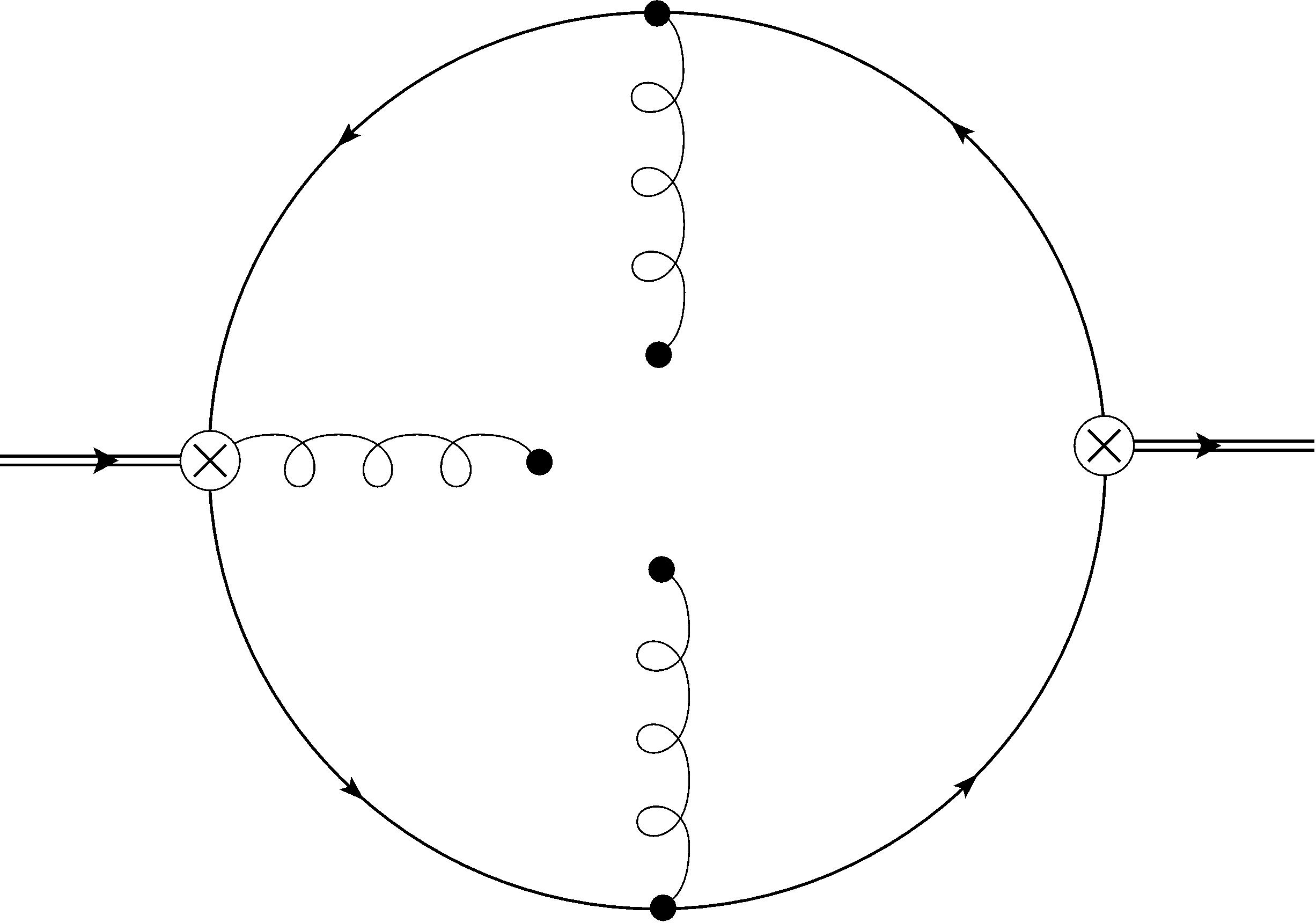} \\
Diagram \rom{4} & Diagram \rom{5}
\end{tabular}
\caption{The Feynman diagrams that contribute to~(\ref{fullCorForm})}
\label{fig01}
\end{figure}

The cross-correlator~(\ref{CorFn}) is calculated within the OPE including
perturbation theory and contributions from the 4d and 6d gluon condensates
defined by
\begin{gather}
   \big\langle \alpha G^{2} \big\rangle = \alpha_{s} \big\langle \! \colon \! G_{\omega\phi}^{a} G_{\omega\phi}^{a} \! \colon\! \big\rangle
    \label{fourDcond}\\
   \big\langle g^{3}G^{3}\big\rangle = g_{s}^{3} f^{abc} \big\langle \! \colon \! G^{a}_{\omega\zeta} \; G^{b}_{\zeta\rho} G^{c}_{\rho\omega} \! \colon\! \big\rangle.
     \label{sixDcond}
\end{gather}
respectively. The corresponding nontrivial diagrams at $\mathcal{O}\big(g_s^3\big)$ are shown in Figure~\ref{fig01}. Wilson coefficients are calculated in the fixed-point gauge~\cite{PascualTarrach1984,BaganAhmadyEliasEtAl1994}. 
We use dimensional regularization in $D=4+2\epsilon$ dimensions
at $\overline{\text{MS}}$ renormalization scale $\mu$. 
We use the following dimensionally regularized 
$\gamma_5$~\cite{AkyeampongDelbourgo1973}
\begin{equation} \label{gamma5conv}
\gamma_{5}=\frac{i}{24}\epsilon_{\mu\nu\sigma\rho}\gamma^{\mu}\gamma^{\nu}\gamma^{\sigma}\gamma^{\rho}.
\end{equation}
We employ TARCER~\cite{MertigScharf1998}, which implements~\cite{Tarasov1996,Tarasov1997}, to express all integrals in terms of a few master integrals each with known solutions~\cite{BoosDavydychev1991,BroadhurstFleischerTarasov1993}.

Denoting the OPE calculation of $\Pi$ by $\Pi^{\text{OPE}}$,
we decompose $\Pi^{\text{OPE}}$ as
\begin{equation} \label{fullCorForm}
  \Pi^{\text{OPE}} = \Pi^{\text{\rom{1}}} + \Pi^{\text{\rom{2}}} + \Pi^{\text{\rom{3}}} + \Pi^{\text{\rom{4}}} + \Pi^{\text{\rom{5}}}
\end{equation}
where the superscripts in~(\ref{fullCorForm}) 
correspond to the labeling scheme of Figure~\ref{fig01}. 
Expanding in $\epsilon$, we find
\begin{gather}
   \Pi^{\text{\rom{1}}}(q^2) =\frac{2\alpha_s m^4 z(1+4z)\:\hypgeom{2}{1}\left(1,1;\frac{5}{2};z\right)}{9\pi^3}\frac{1}{\epsilon}+\cdots
    \label{exppert}\\
   \Pi^{\text{\rom{2}}}(q^2) = \frac{3z - z \;\hypgeom{2}{1}\big(1,1;\frac{5}{2};z\big)}{18\pi(1-z)}
\big\langle\alpha G^{2} \big\rangle
     \label{expfourd}\\
\begin{aligned}
\Pi^{\text{\rom{3}}}(q^2) &= \frac{\big\langle g^{3} G^{3} \big\rangle}{1152\pi^2 m^2 (1-z)^3}\Bigg(-2-5z+4z^2 \\
&\;\;+(2-7z+10z^2-4z^3)\:\hypgeom{2}{1}\big(1,1;\frac{5}{2};z\big)\Bigg)
\end{aligned}
     \label{expsixd}\\
\begin{aligned}
\Pi^{\text{\rom{4}}}(q^2) &= \frac{\big\langle g^{3} G^{3} \big\rangle}{4608\pi^2 m^2 (1-z)^3}\Bigg( 22-41z+16z^2 \\
&\!\!\!-(10-25z+22z^2-8z^3)\:\hypgeom{2}{1}\big(1,1;\frac{5}{2};z\big)\Bigg)
\end{aligned}
    \label{expsixdA}\\
\begin{aligned}
\Pi^{\text{\rom{5}}}(q^2) &= \frac{\big\langle g^{3} G^{3} \big\rangle}{1536\pi^2 m^2 (1-z)^2}\Bigg(-15+12z\\
&\;\;\;\;\;\;\;\;\;\;\;\;\;\;\;\;\;+(3-2z)\:\hypgeom{2}{1}\big(1,1;\frac{5}{2};z\big)\Bigg)
\end{aligned}
     \label{expsixdB}
\end{gather}
where $m$ is a heavy quark mass, $z=q^2/(4m^2)$
and the $\hypgeom{p}{q}(\cdots;\cdots;z)$ are generalized hypergeometric 
functions~\cite{AbramowitzStegun1965}. In~(\ref{exppert}) we include only a divergent term.

\begin{figure}[ht]
\begin{tabular}{cc}
  \includegraphics[width=35mm]{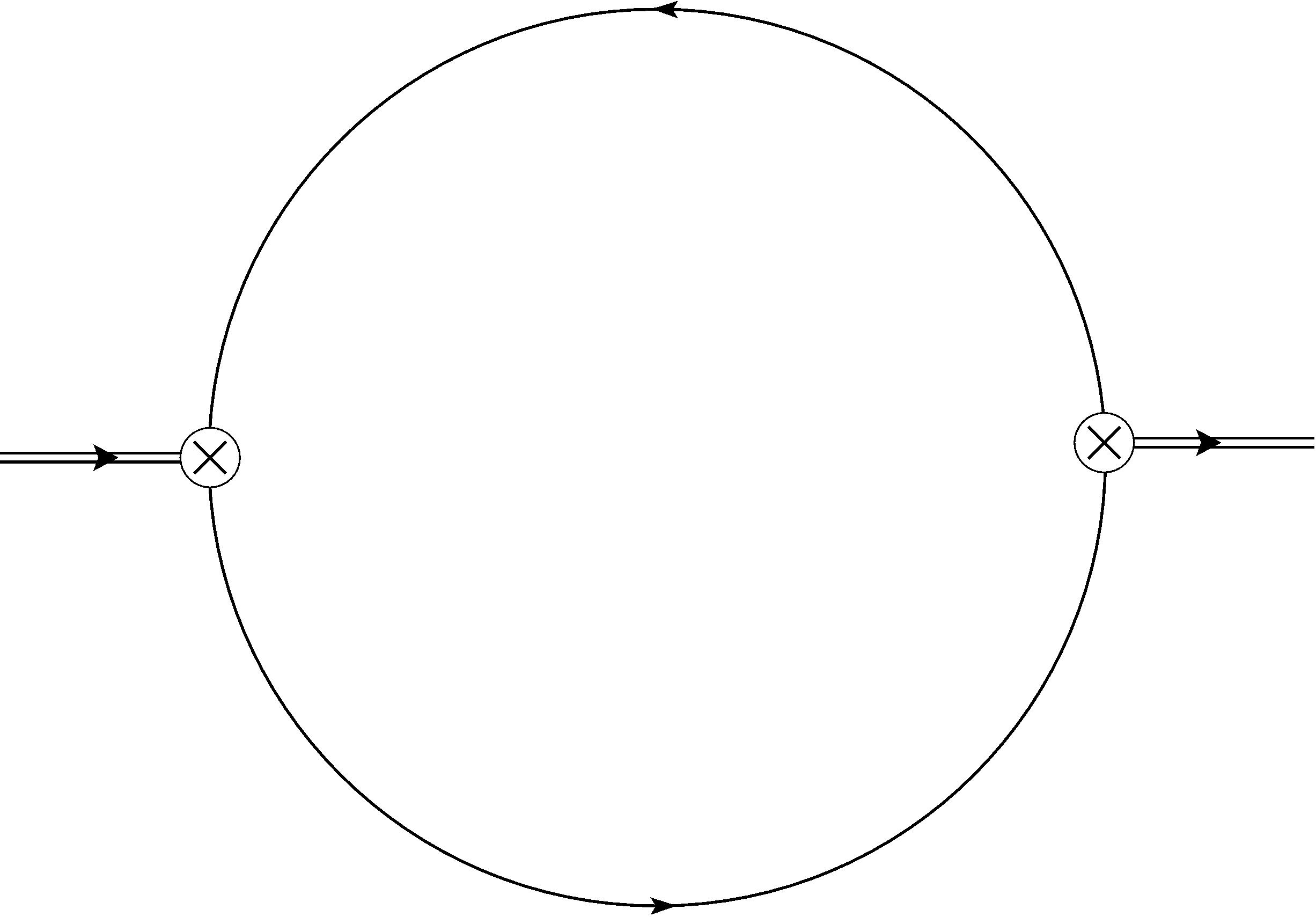} &   \includegraphics[width=35mm]{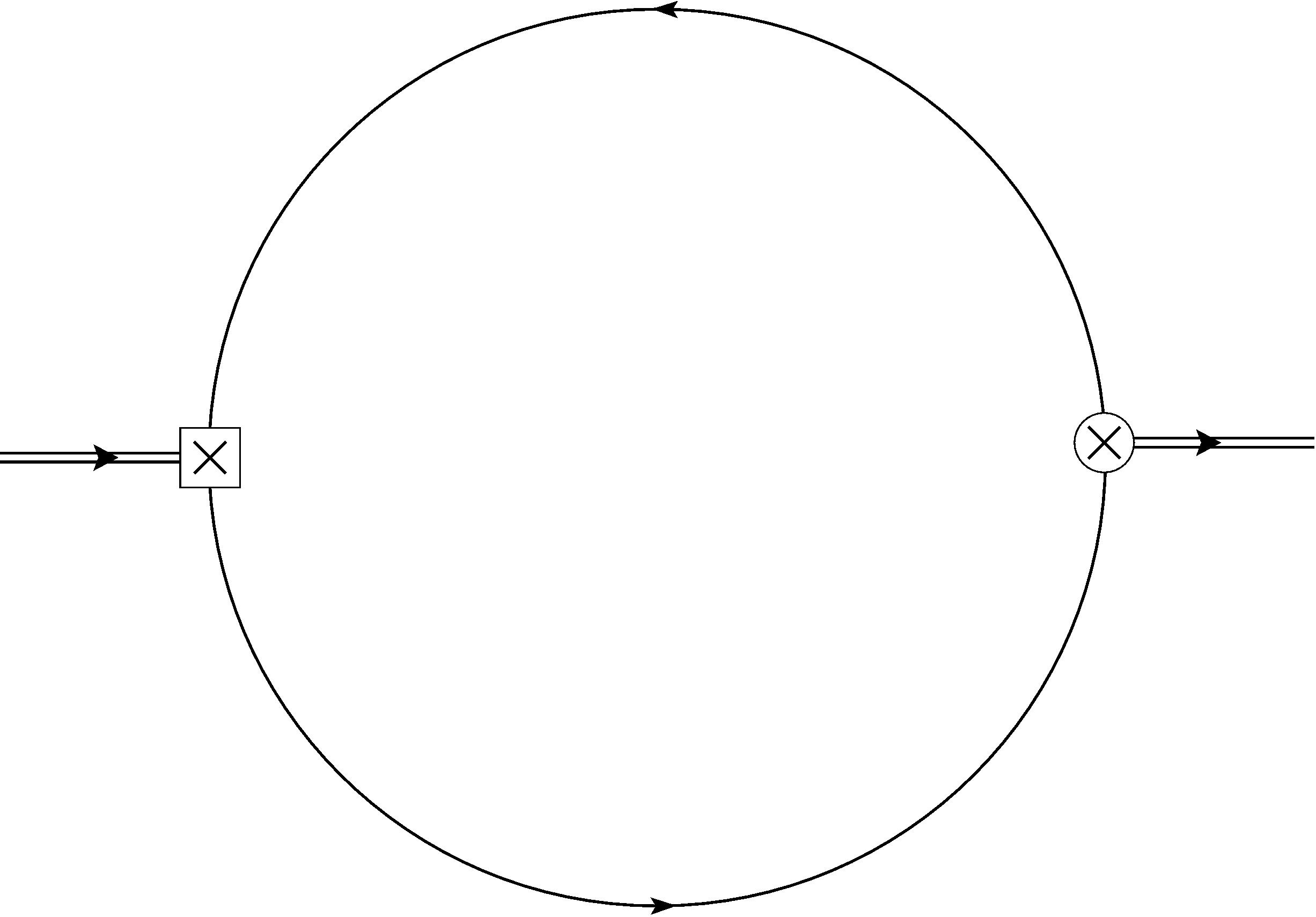} \\
\end{tabular}
\caption{Renormalization-induced diagrams generated by~(\ref{renormTheLast}), the square insertion represents current~(\ref{DCurrent}).}
\label{fig02}
\end{figure}

The perturbative result~(\ref{exppert}) contains a non-polynomial divergence which we eliminate via operator mixing under renormalization~\cite{Chen:2013pya,ho:2017}. The mesonic current~(\ref{CurMes}) is renormalization-group (RG) invariant;
thus, only operator mixing of the hybrid current~(\ref{CurHyb}) needs to be considered. Replacing the bare hybrid current~(\ref{CurHyb}) in~(\ref{CorFn}) by
\begin{equation} \label{renormTheLast}
  j_{\nu}^{(\text{h})}  \rightarrow  j_{\nu}^{(\text{h})} 
  + \frac{C_{1}}{\epsilon}j_{\nu}^{(\text{m})} 
  + \frac{C_{2}}{\epsilon}j_{\nu}^{(\text{c})}
\end{equation}
where
\begin{equation} \label{DCurrent}
j_{\nu}^{(c)} = \overline{Q} i D_{\nu} Q
\end{equation}
%
%
%
and where $C_1$ and $C_2$ are as-yet-undetermined renormalization constants generates two renormalization-induced diagrams shown in Figure~\ref{fig02}.
Evaluating these diagrams and choosing $C_1$ and $C_2$ such that the total perturbative contribution is divergences free, we find
\begin{equation}
C_{1} = -\frac{10 m^2 \alpha_s}{9 \pi} \;\;\;\;\;\;\;\;\; C_{2} = \frac{4 m \alpha_s}{9 \pi}.
\end{equation}
The updated perturbative result that replaces~(\ref{exppert}) is
\begin{equation} \label{expRenormedPert}
\begin{aligned} 
  \Pi^{(\text{\rom{1}})}(q^2) = &\frac{2 \alpha_s m^4 z}{81 \pi^3} 
  \Bigg(18(z-1)\:\hypgeom{3}{2}\big(1,1,1;\tfrac{3}{2},3;z\big) \\
  & -2z (4 z + 1)\:\hypgeom{3}{2}\big(1,1,2;\tfrac{5}{2},4;z\big)\\
  & +3 \Big(3 (4 z+1) \log \left(\frac{m^2}{\mu ^2}\right)+26 z+6 \Big)\\
  & \;\;\;\;\;\;\; \times \:\hypgeom{2}{1}\big(1,1;\tfrac{5}{2};z\big) \Bigg).
\end{aligned}
\end{equation}

In summary, $\Pi^{\text{OPE}}$ is the sum of~(\ref{expRenormedPert}) 
and~(\ref{expfourd}) -- (\ref{expsixdB}).

\section{QCD Laplace Sum-Rules}

The correlator $\Pi$ defined in~(\ref{CorFnProj}) satisfies the dispersion relation
\begin{equation}\label{dispersion_relation}
  \Pi(Q^2)=\frac{Q^6}{\pi}\int_{t_0}^{\infty}
  \frac{\mathrm{Im}\Pi(t)}{t^3(t+Q^2)}
  dt +\cdots
\end{equation}
where $Q^2=-q^2>0$. 
On the left-hand side, we let $\Pi \to \Pi^{\text{OPE}}$;
on the right-hand side, we identify $\mathrm{Im}\Pi(t)$ as the hadronic spectral function. To eliminate the subtraction constants in~(\ref{dispersion_relation}) represented by $\cdots$ and to accentuate the resonance contributions of the spectral function, we apply the Borel transform 
\begin{equation}\label{borel}
  \hat{\mathcal{B}}=\!\lim_{\stackrel{N,Q^2\rightarrow\infty}{\tau=N/Q^2}}
  \!\frac{\big(-Q^2\big)^N}{\Gamma(N)}\bigg(\frac{d}{dQ^2}\bigg)^N
\end{equation}
with Borel parameter $\tau$, and form the $0^{\text{th}}$-order LSR~\cite{Shifman:1978bx} 
\begin{equation} \label{zeroThLSR}
\lsr(\tau)\equiv\frac{1}{\tau}\hat{\mathcal{B}}\Big\{\Pi(Q^2) \Big\}\\
= \int_{4m^2}^{\infty} e^{-t\tau}\frac{1}{\pi}\mathrm{Im}\Pi(t)dt.
\end{equation}
We assume a resonance(s) plus continuum model
\begin{equation} \label{ResCont}
\frac{1}{\pi}\mathrm{Im}\Pi(t)
    =\rho^{\text{(had)}}(t)+\frac{1}{\pi}\mathrm{Im}\Pi^{\text{OPE}}(t)\theta(t-s_0)
\end{equation}
where $\rho^{\text{(had)}}$ represents the resonance portion of the spectral function and $s_0$ is the continuum threshold. (We discuss $\rho^{\text{(had)}}$ in Section~\ref{IV}.) 
Then, we define the continuum-subtracted $0^{\text{th}}$-order LSR as
\begin{equation} \label{subedLSR}
\begin{aligned} 
\lsr(\tau, s_0) & \equiv \lsr(\tau) - \int_{s_0}^{\infty} e^{-t\tau}\frac{1}{\pi}\mathrm{Im}\Pi^{\text{OPE}}(t)dt \\
& =\int_{t_0}^{s_0} \rho^{\text{(had)}}(t)dt
\end{aligned} 
\end{equation}
where $t_0$ is the hadronic threshold.

\begin{figure}[ht]
\includegraphics[width=80mm]{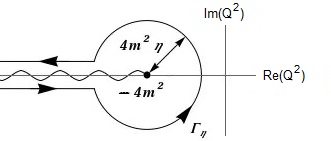}
\caption{Integration contour used in the computation of the LSR~(\ref{intermediateTheoryLSR})}
\label{fig03}
\end{figure}

To evaluate~(\ref{subedLSR}), 
we use the following relationship between the Borel transform and the inverse Laplace transform~\cite{Shifman:1978bx}
\begin{equation}\label{borelIdentity}
\begin{aligned}
  \frac{1}{\tau}\hat{\mathcal{B}} \Big\{ f(Q^2) \Big\}&=\hat{\mathcal{L}}^{-1} \Big\{ f(Q^2) \Big\}\\
  &= \frac{1}{2 \pi i} \int_{c-i \infty}^{c+i \infty} f(Q^2) e^{Q^2 \tau} dQ^2
\end{aligned}
\end{equation}
where $c$ is a real constant for which $f(Q^2)$ is analytic for 
$\text{Re}(Q^2)>c$.
The function $\Pi^{\text{OPE}}$ has a branch cut along the real axis for $Q^2<-4m^2$. 
Letting $f(Q^2) \to \Pi^{\text{OPE}}(Q^2)$ in~(\ref{borelIdentity})
and warping the integration contour to that shown in Figure~\ref{fig03},
we can write
\begin{equation}\label{intermediateTheoryLSR}
\begin{aligned}
  \lsr\double{\tau}{s_0}=& \int_{4m^2(1+\eta)}^{s_0}e^{-t\tau}\frac{1}{\pi}
    \text{Im}\Pi^{\text{OPE}}(t) \, dt\\
  &+\frac{1}{2\pi i}\int_{\Gamma_{\eta}} e^{Q^2 \tau}
    \Pi^{\text{OPE}}(Q^2) \, dQ^2
\end{aligned}
\end{equation}
for $\eta \to 0^{+}$.
%
%
%

RG improvement~\cite{Narison:1981ts} requires that the strong coupling and quark mass in the LSR~(\ref{intermediateTheoryLSR})
get replaced by running quantities evaluated at renormalization scale $\mu$. At one-loop in $\overline{\text{MS}}$, we make the following replacements: 
for charmonium, we set $\mu \to \overline{m}_c$ in
\begin{gather}
  \alpha_s(\mu) = \frac{\alpha_s(M_{\tau})}{1 + \frac{25 \alpha_s(M_{\tau})}%
  {12\pi}\log\!{\Big(\frac{\mu^2}{M_{\tau}^2}\Big)}}
  \\ 
  m_{c}(\mu) = \overline{m}_{c}\bigg(\frac{\alpha_s(\mu)}
  {\alpha_s(\overline{m}_{c})}\bigg)^{12/25}
\end{gather}
whereas, for bottomonium, we set $\mu \to \overline{m}_b$ in
\begin{gather}
  \alpha_s(\mu) = \frac{\alpha_s(M_Z)}{1 + \frac{23 \alpha_s(M_Z)}%
  {12\pi}\log\!{\Big(\frac{\mu^2}{M_Z^2}\Big)}}
  \\ 
  m_{b}(\mu) = \overline{m}_{b}\bigg(\frac{\alpha_s(\mu)}
  {\alpha_s(\overline{m}_{b})}\bigg)^{12/23}
\end{gather}
where~\cite{Olive:2016xmw}
\begin{gather}
  \alpha_s(M_{\tau})=0.330\pm0.014\label{alphatau}\\
  \alpha_s(M_{Z})=0.1185\pm0.0006\label{alphaZ}\\
  \overline{m}_c=(1.275\pm0.025)\ \text{GeV}\label{charmMass}\\
  \overline{m}_b=(4.18\pm 0.03)\ \text{GeV}\label{bottomMass}.
\end{gather}
The condensate values used are~\cite{Launer:1983ib,Narison2010}
\begin{gather}
  \glueFourD=(0.075\pm0.02)\ \text{GeV}^4\label{glueFourDValue}\\
  \glueSixD=(8.2\pm1.0)\ \text{GeV}^2)\label{glueSixDValue} \, \glueFourD.
\end{gather}

\section{Analysis}\label{IV}

We select a Borel window $\double{\tau_{\text{min}}}{\tau_{\text{max}}}$ over which we will examine the LSR using the methodology of~\cite{Chen:2013pya,ho:2017,BergHarnettKleivEtAl2012,Harnett:2012gs}. 
We require that the LSR converges in the sense that the perturbative contribution is at least three times that of the 4d gluon condensate contribution which, in turn, is at least three times that of the 6d gluon condensate contribution. 
We also require that the resonance(s) contribution to the LSR is at least 10\% that of the total resonance(s) plus continuum contributions i.e.,
\begin{equation}
  \frac{\lsr\double{\tau}{s_0}}{\lsr\double{\tau}{\infty}} > 10\%.
\end{equation}
We find that the Borel window for charmonium is
$0.1 \, \text{GeV}^{-2}\leq\tau\leq0.6 \, \text{GeV}^{-2}$ 
and for bottomonium is $0.01 \, \text{GeV}^{-2}\leq\tau\leq0.2 \, \text{GeV}^{-2}$.

The quantity
$\rho^{\text{(had)}}$ in~(\ref{ResCont}) will contain resonances that are to be tested for coupling to both the meson~(\ref{CurMes}) and hybrid~(\ref{CurHyb}) currents. 
The vector charmonium and bottomonium resonances listed by the 
Particle Data Group~\cite{Olive:2016xmw} are summarized in Table~\ref{mspectrum}. 

\begin{table}[ht]
\caption{Heavy vector quarkonium resonances.}
\label{mspectrum}
\begin{tabular}{cc}
\hline
    \begin{minipage}{.43\linewidth}
    \scalebox{0.92}{
        \begin{tabular}{lc}
          \multicolumn{2}{c}{Charmonium} \\
          \hline
       	  Name & Mass (GeV)\\
          \hline
          $J/\psi$ & 3.10 \\
		  $\psi(2S)$ & 3.69 \\
		  $\psi(3770)$ & 3.77 \\
		  $\psi(4040)$ & 4.04 \\
		  $\psi(4160)$ & 4.19 \\
		  $X(4230)$ & 4.23 \\
		  $X(4260)$ & 4.23 \\
		  $X(4360)$ & 4.34 \\
		  $\psi(4415)$ & 4.42 \\
		  $X(4660)$ & 4.64 \\
		\end{tabular}}
    \end{minipage} &

    \begin{minipage}{.43\linewidth}
    \scalebox{0.92}{
        \begin{tabular}{lc}
          \multicolumn{2}{c}{Bottomonium} \\
          \hline
       	  Name & Mass (GeV)\\
          \hline
		  $\Upsilon(1S)$ & 9.46 \\
		  $\Upsilon(2S)$ & 10.02 \\
		  $\Upsilon(3S)$ &  10.34 \\
		  $\Upsilon(4S)$ &  10.58 \\
		  $\Upsilon(10860)$ & 10.89 \\
		  $\Upsilon(11020)$ & 10.99 \\
        \end{tabular}}
    \end{minipage}\\
\hline
\multicolumn{2}{@{}p{75mm}}{
\footnotesize{Note: in the charmonium sector entries labeled with $X$ are those with unknown $I^G$ whereas $\psi$ indicates $I^G=0^{-}$.}
                             }
\end{tabular}
\end{table}

All resonances in Table~\ref{mspectrum} have widths of $\lesssim100$~MeV. As LSRs are quite insensitive to resonance width, we ignore the widths of individual resonances. For collections of resonances with masses separated by 250~MeV or less, we cluster these resonances into a single resonance with nonzero (or, in some models, zero) width. 
We consider a variety of $\rho^{\text{(had)}}$ of the form
%
%
%
\begin{equation}\label{rhoForms}
\begin{aligned}
  \rho^{\text{(had)}}(t)=\sum_{i=1}^n
  \begin{cases}
    \xi_i \delta(t-m_i^2),\ \Gamma_i=0 \\
    \frac{\xi_i}{2m_i \Gamma_i} \theta \big(t-m_i(m_i-\Gamma_i)\big)\\
    \;\;\; \times \; \theta\big(m_i(m_i+\Gamma_i)-t\big),\ \Gamma_i\neq0
  \end{cases}
\end{aligned}
\end{equation}
where $n$ is the number of distinct resonances or resonance clusters 
and $\Gamma_i$ are the resonance cluster widths.
The quantities $\xi_i$ are mixing parameters. Resonances with both meson and hybrid components (i.e., mixed states) have $\xi_i\neq0$; pure meson or pure hybrid states have $\xi_i=0$. Substituting~(\ref{rhoForms}) into~(\ref{subedLSR}) gives
%
%
%
\begin{equation}\label{specific_rho}
  \lsr\double{\tau}{s_0}=\sum_{i=1}^n
  \begin{cases}
    \xi_i e^{-m_i^2 \tau},\ \Gamma_i=0\\
    \xi_i e^{-m_i^2 \tau}\frac{\sinh\big(m_i \Gamma_i\tau\big)}{m_i\Gamma_i\tau},\ 
      \Gamma_i\neq0
  \end{cases} \!\!\!\! .
\end{equation}

Informed by Table~\ref{mspectrum}, we input 
a variety of choices for $m_i$ and $\Gamma_i$ into~(\ref{specific_rho}). 
We then partition the Borel window $\double{\tau_{\text{min}}}{\tau_{\text{max}}}$ into $N=20$ equal length subintervals 
using grid points $\{\tau_j\}_{j=0}^N$ and define
\begin{equation}\label{chiSquaredGeneral}
\begin{aligned}
  \chi^2(\xi_1,&\ldots,\xi_n,\,s_0)= \\ 
  & \!\! \sum_{j=0}^N
  \Bigg(
    \lsr\double{\tau_j}{s_0}-
    \sum_{i=1}^n \int_{4m^2}^{s_0} e^{-t\tau_j}
    \rho^{\text{(had)}}_i(t) dt
  \Bigg)^2.
\end{aligned}
\end{equation}
By minimizing~(\ref{chiSquaredGeneral}), we extract $\xi_i$ and $s_0$ as best-fit parameters. 
Table~\ref{charmoniumResults} contains the models used and results obtained in the charmonium sector, 
and Table~\ref{bottomoniumResults} contains similar information for the bottomonium sector. Rather than present each $\xi_i$, we instead present $\zeta$ and $\frac{\xi_i}{\zeta}$ where
\begin{equation}\label{zeta}
  \zeta=\sum_{i=1}^n |\xi_i|.
\end{equation}
The errors in Tables~\ref{charmoniumResults} and~\ref{bottomoniumResults} include a variation of~$\pm 0.1$~GeV in the renormalization scale $\mu$. We also vary the end points of the Borel window by $\pm 0.05\ \text{GeV}^{-2}$ in the charmonium analysis and by $\pm 0.005\ \text{GeV}^{-2}$ in the bottomonium analysis. 
Additionally, we include the errors associated with the quantities~(\ref{alphatau}) -- (\ref{glueSixDValue}). Our results are most sensitive to uncertainties in the quark mass parameters~(\ref{charmMass}) and~(\ref{bottomMass}). 

\section{Discussion}\label{V}

Looking at the $\chi^2$ values in Tables~\ref{charmoniumResults} and~\ref{bottomoniumResults}, we see that the inclusion of a third heavy resonance cluster in our models significantly improves the fits. Note that these third resonance clusters make large contributions to the LSR in spite of the LSR's suppression of higher mass resonance contributions. Consider a quantitative measure of the third resonance's contribution given by
\begin{equation}\label{signalStrength}
  \frac{\int_{t_0}^{s_0} e^{-t\tau}\rho_3^{(\text{had})}(t)dt}%
  {\sum_{i=1}^{3}\left|\int_{t_0}^{s_0} e^{-t\tau}\rho_i^{(\text{had})}(t)dt\right|}.
\end{equation}
For example, evaluating~(\ref{signalStrength}) for Model C3 of Table~\ref{charmoniumResults} gives $0.43$. In the bottomonium sector, Model B3 of Table~\ref{bottomoniumResults} gives $0.35$. The $\chi^2$ values also indicate that the inclusion of resonance widths (i.e., $\Gamma_i \neq 0$) has no significant impact on the fits. This insensitivity to resonance width is a characteristic of LSRs and is expected.

Models including a fourth resonance were also examined, but, in all cases, led to $\chi^2$ minima occurring at $s_0 \! \approx \! m_4^2$. 
In other words, these fourth resonances merged with the continuum in contradiction with the separation between resonances and continuum assumed 
in~(\ref{ResCont}).  Furthermore, the two resonance models examined in Tables~\ref{charmoniumResults} and~\ref{bottomoniumResults} (i.e., Models C2 and B2) suffer from this same problem making them too unreliable.

Examining the three resonance models in the charmonium sector (Models C3, C4 and C5), we find a nonzero mixing parameter for the $J/\psi$; no evidence for mixing in the $\psi(2S),\,\psi(3770)$ cluster; and a large mixing parameter for the $4.3$~GeV cluster. This $4.3$~GeV cluster represents a grouping of all resonances from the $\psi(4040)$ to the $X(4660)$. The effects of varying $m_3$ in this system from 4.0~GeV--4.6~GeV were investigated, and we found that $m_3=4.3$~GeV yielded the smallest minimum $\chi^2$ value. As the $\psi(2S),\,\psi(3770)$ cluster exhibits no mixing, the final two models in Table~\ref{charmoniumResults} (Models C6 and C7) exclude this cluster; doing so has minimal effect on the values of $\xi_1,\ \xi_3, s_0$ and the minimum value of $\chi^2$.

In the bottomonium sector, the three resonance models in Table~\ref{bottomoniumResults} (Models B3 and B4) show a nonzero mixing parameter for all three resonances. Thus, the $\Upsilon(1S)$, $\Upsilon(2S)$, and $\Upsilon(3S),\,\Upsilon(4S)$ cluster all exhibit mixing.

To summarize, in both the charmonium and bottomonium sectors, we find that the addition of a third heavy resonance cluster improves the agreement between theory and experiment significantly. In the charmonium sector, meson-hybrid mixing is well-described by a two resonance model consisting of the $J/\psi$ and a second state with mass $\approx4.3$~GeV. This result is consistent with the hypothesis that the $X(4260)$ could be a resonance with significant hybrid content~\cite{ClosePage1995a,Kou:2005gt,Zhu:2005hp}. In the bottomonium sector, our results indicate nonzero mixing in the $\Upsilon(1S)$, $\Upsilon(2S)$, and $\Upsilon(3S),\,\Upsilon(4S)$ cluster.\\

\textbf{Acknowledgements-} We are grateful for financial support from the Natural Sciences and Engineering
Research Council of Canada (NSERC).

\begin{center}
\begin{table*}[ht]
\centering
\caption{Models and results in the charmonium sector}
\label{charmoniumResults}
\scalebox{0.85}{
\begin{tabular}{c l l l c c c c c c}
\hline
Model & $m_1$ , $\Gamma_1$  & $m_2$ , $\Gamma_2$  & $m_3$ , $\Gamma_3$ & $s_0$ & $\chi^2\times10^6$ & $\zeta$ & \multirow{2}{*}{$\frac{\xi_1}{\zeta}$} & \multirow{2}{*}{$\frac{\xi_2}{\zeta}$} & \multirow{2}{*}{$\frac{\xi_3}{\zeta}$} \\
\# & \scriptsize{(GeV)}  & \scriptsize{(GeV)} & \scriptsize{(GeV)} & \scriptsize{(GeV$^2$)} & \scriptsize{(GeV$^{12}$)} & \scriptsize{(GeV$^6$)} & & &\\
\hline
C1 & 3.10 , 0 & - & - & 12.4 & 3.32 & $0.507 \pm 0.022$ & 1 & $-$  & $-$ \\
C2 & 3.10 , 0 & 3.73 , 0 & - & 13.9 & 2.23 & $0.736 \pm 0.043$ & $0.716 \pm 0.035$ & $ \;\;\, 0.284 \pm 0.035$ & $-$ \\
C3 & 3.10 , 0 & 3.73 , 0 & 4.30 , 0 & 23.7 & 0.105 & $2.679 \pm 0.254$ & $0.224 \pm 0.014$ & $0.010 \pm 0.055$ & $0.766 \pm 0.042$ \\
C4 & 3.10 , 0 & 3.73 , 0 & 4.30 , 0.30 & 23.8 & 0.103 & $2.709 \pm 0.274$ & $0.222 \pm 0.015$ & $-0.001 \pm 0.054$ & $0.777 \pm 0.044$ \\
C5 & 3.10 , 0 & 3.73 , 0.05 & 4.30 , 0.30 & 23.8 & 0.103 & $2.709 \pm 0.274$ & $0.222 \pm 0.015$ & $-0.001 \pm 0.054$ & $0.777 \pm 0.044$ \\
C6 & 3.10 , 0 & - & 4.30 , 0 & 23.9 & 0.109 & $2.713 \pm 0.263$ & $0.223 \pm 0.019$ & $-$ & $0.777 \pm 0.019$ \\
C7 & 3.10 , 0 & - & 4.30 , 0.30 & 23.8 & 0.103 & $2.699 \pm 0.262$ & $0.223 \pm 0.020$ & $-$ & $0.777 \pm 0.020$ \\
\hline
\end{tabular}
}
\end{table*}
\end{center}
\begin{center}
\begin{table*}[ht]
\centering
\caption{Models and results in the bottomonium sector}
\label{bottomoniumResults}
\scalebox{0.85}{
\begin{tabular}{c l l l c c c c c c}
\hline
Model & $m_1$ , $\Gamma_1$  & $m_2$ , $\Gamma_2$  & $m_3$ , $\Gamma_3$ & $s_0$ & $\chi^2\times10^4$ & $\zeta$ & \multirow{2}{*}{$\frac{\xi_1}{\zeta}$} & \multirow{2}{*}{$\frac{\xi_2}{\zeta}$} & \multirow{2}{*}{$\frac{\xi_3}{\zeta}$} \\
\# & \scriptsize{(GeV)}  & \scriptsize{(GeV)} & \scriptsize{(GeV)} & \scriptsize{(GeV$^2$)} & \scriptsize{(GeV$^{12}$)} & \scriptsize{(GeV$^6$)} & & &\\
\hline
 B1 & 9.46 , 0 & - & - & 105 & 169 & $132 \pm 4$ & 1 & $-$ & $-$ \\
 B2 & 9.46 , 0 & 10.02 , 0 & - & 100 & 154 & $165 \pm 8$ & $0.824 \pm 0.015$ & $-0.176 \pm 0.015$ & $-$ \\
 B3 & 9.46 , 0 & 10.02 , 0 & 10.47 , 0 & 130 & 0.18 & $1255 \pm 83$ & $0.213 \pm 0.004$ & $-0.374 \pm 0.006$ & $0.413 \pm 0.004$ \\
 B4 & 9.46 , 0 & 10.02 , 0 & 10.47 , 0.22 & 130 & 0.18 & $1259 \pm 83$ & $0.212 \pm 0.004$ & $-0.375 \pm 0.006$ & $0.413 \pm 0.004$ \\
\hline
\end{tabular}
}
\end{table*}
\end{center}

\vspace*{-20mm}

\bibliographystyle{elsarticle-num}
\bibliography{research}

\end{document}